\documentstyle[prd,aps,12pt,psfig]{revtex}
\begin{document}
\newcommand{\CVT}{ {\scriptscriptstyle {CVT}} }
\newcommand{\A}{ {\scriptscriptstyle {A}} }
\newcommand{\oned}{ {\scriptscriptstyle {1D}} }
\newcommand{\threed}{ {\scriptscriptstyle {3D}} }
\newcommand{\mb}[1]{\mbox{\boldmath $#1$}}
\newcommand{\gtsima}{$\; \buildrel > \over \sim \;$}
\newcommand{\ltsima}{$\; \buildrel < \over \sim \;$}
\newcommand{\simgt}{\lower.5ex\hbox{\gtsima}}
\newcommand{\simlt}{\lower.5ex\hbox{\ltsima}}
\newcommand{\hikpc}{{\hbox {$h^{-1}$}{\rm kpc}} }
\newcommand{\himpc}{{\hbox {$h^{-1}$}{\rm Mpc}} }
\newcommand{\0}{ {\scriptscriptstyle {0}} }
\newcommand{\T}{ {\scriptscriptstyle {\rm T}} }
\newcommand{\kms}{ {\rm km/sec} }
\newcommand{\keV}{ {\rm keV} }
\newcommand{\mpc}{ {\rm Mpc} }
\newcommand{\bfs}{{\mbox{\boldmath $s$}}}
\newcommand{\bfx}{{\mbox{\boldmath $x$}}}
\newcommand{\bfk}{{\mbox{\boldmath $k$}}}
\newcommand{\smbfk}{{\mbox{\footnotesize\boldmath $k$}}}
\newcommand{\smbfx}{{\mbox{\footnotesize\boldmath $x$}}}
\newcommand{\sVert}{{\scriptscriptstyle\Vert}}
\def\pp{\par\parshape 2 0truecm 15.5truecm 1truecm 14.5truecm\noindent}

\vspace*{-1.5cm}
\leftline{\small To appear in the 
 Proceedings of Asia Pacific Center for Theoretical Physics 
 Inaugration Conference.}
\vspace*{-0.2cm}
\rightline{UTAP-235/96, RESCEU-22/96}

\begin{center}
{\large\bf $\Omega_0$ and $\lambda_0$ from galaxy and quasar
  clustering:\\ cosmic virial theorem and cosmological redshift-space
  distortion}

Yasushi Suto
\end{center}

\begin{center}
\baselineskip=15pt
{\sl Department of Physics and 
RESCEU  (Research Center for the Early Universe)\\
 School of  Science, The University of Tokyo, Tokyo 113, Japan.}
\end{center}

\begin{abstract}
\baselineskip=15pt
  I discuss two cosmological tests to determine the cosmological
  density parameter $\Omega_0$ the cosmological constant $\lambda_0$,
  which make use of the anisotropy of the two-point correlation
  functions due to the peculiar velocity field and the cosmological
  redshift-space distortion.
\end{abstract}


\section{Introduction}

The three-dimensional distribution of galaxies in the redshift surveys
differ from the true one since the distance to each galaxy cannot be
determined by its redshift $z$ only; for $z \ll 1$ the peculiar
velocity of galaxies, typically $\sim (100-1000)\kms$, contaminates
the true recession velocity of the Hubble flow, while the true
distance for objects at $z\simgt1$ sensitively depends on the (unknown
and thus assumed) cosmological parameters. This hampers the effort to
understand the true distribution of large-scale structure of the
universe. Nevertheless such redshift-space distortion effects are
quite useful since through the detailed theoretical modelling, one can
derive the peculiar velocity dispersions of galaxies as a function of
separation, and also can infer the cosmological density parameter
$\Omega_0$ and the dimensionless cosmological constant $\lambda_0$,
for instance. In what follows, I will present two specific topics
concerning the redshift distortion; the small-scale pair-wise peculiar
velocity dispersions of galaxies\cite{suto93}, and anisotropies in the
two-point correlation functions\cite{ms96} at high redshifts.

\section{$\Omega_0$ from nonlinear galaxy clustering}

Assuming that particle pairs in expanding universes are in
``statistical equilibrium'' on small scales, their relative peculiar
velocity dispersion can be computed as a function of their separation.
The result is called the cosmic virial theorem \cite{peebcvt,lss}
(CVT, hereafter) which predicts the one-dimensional pair-wise relative
peculiar velocity dispersion $\sigma_{\oned, \CVT}$ as a function of
$\Omega_0$. 

The observed two- and three-point correlation functions of {\it
  galaxies}, $\xi_{\rm g}$ and $\zeta_{\rm g}$, are well approximated
by the simple form\cite{gp77,dp83}:
\begin{eqnarray}
  \xi_{\rm g}(r) = \left( {r_\0 \over r}\right)^\gamma , \qquad
  \zeta_{\rm g} (r_1, r_2, r_3) = Q_{\rm g} \, [\xi_{\rm
    g}(r_1)\xi_{\rm g}(r_2) + \xi_{\rm g}(r_2)\xi_{\rm g}(r_3) +
  \xi_{\rm g}(r_3)\xi_{\rm g}(r_1)]
\end{eqnarray}
where $r_0=(5.4\pm0.3)\himpc$, $\gamma=1.77\pm0.04$, $Q_{\rm
  g}=1.29\pm 0.21$. Thus it is reasonable to assume that the two- and
three-point correlation functions of {\it mass}, $\xi_\rho$ and
$\zeta_\rho$, also obey the same scaling except for the overall
amplitude:
\begin{eqnarray}
  \xi_\rho(r) = {1 \over b_{\rm g}^2} \xi_{\rm g}(r), \qquad
  \zeta_\rho (r_1, r_2, r_3) = {Q_\rho  \over Q_{\rm g} b_{\rm g}^4}
  \zeta_{\rm g} (r_1, r_2, r_3) .
\label{eq:hier}
\end{eqnarray}
Then the CVT prediction of the small-scale peculiar velocity
dispersion is given by\cite{suto93,peebcvt}
\begin{eqnarray}
\label{eq:cvt1d}
 && \sigma_{\oned, \CVT}(r) = 1460
     \sqrt{\Omega_0 Q_\rho \over 1.3 b_{\rm g}^2}
  \sqrt{I(\gamma) \over 33.2} 5.4^{\frac {\gamma-1.8}2} \left({r_\0
      \over 5.4\himpc} \right)^{\frac\gamma 2} \left({r \over 1\himpc }
  \right)^{\frac{2-\gamma}2} \kms , \\
&&I(\gamma) \equiv {\pi \over (\gamma-1)(2-\gamma)(4-\gamma)} \nonumber
\\
&& ~~~~\times \int_0^\infty \, dx\, {1+x^{-\gamma} \over x^2} 
\left\{ 
(1+x)^{4-\gamma}-|1+x|^{4-\gamma}
-(4-\gamma)x\left[(1+x)^{2-\gamma}+|1+x|^{2-\gamma}\right]
\right\} ,
\end{eqnarray}
and numerically $I(1.65)\sim 25.4$, $I(1.8)\sim 33.2$, and 
$I(1.95)\sim 55.6$.

\begin{figure}
\vspace{-1.0cm}
\begin{center}
   \leavevmode\psfig{figure=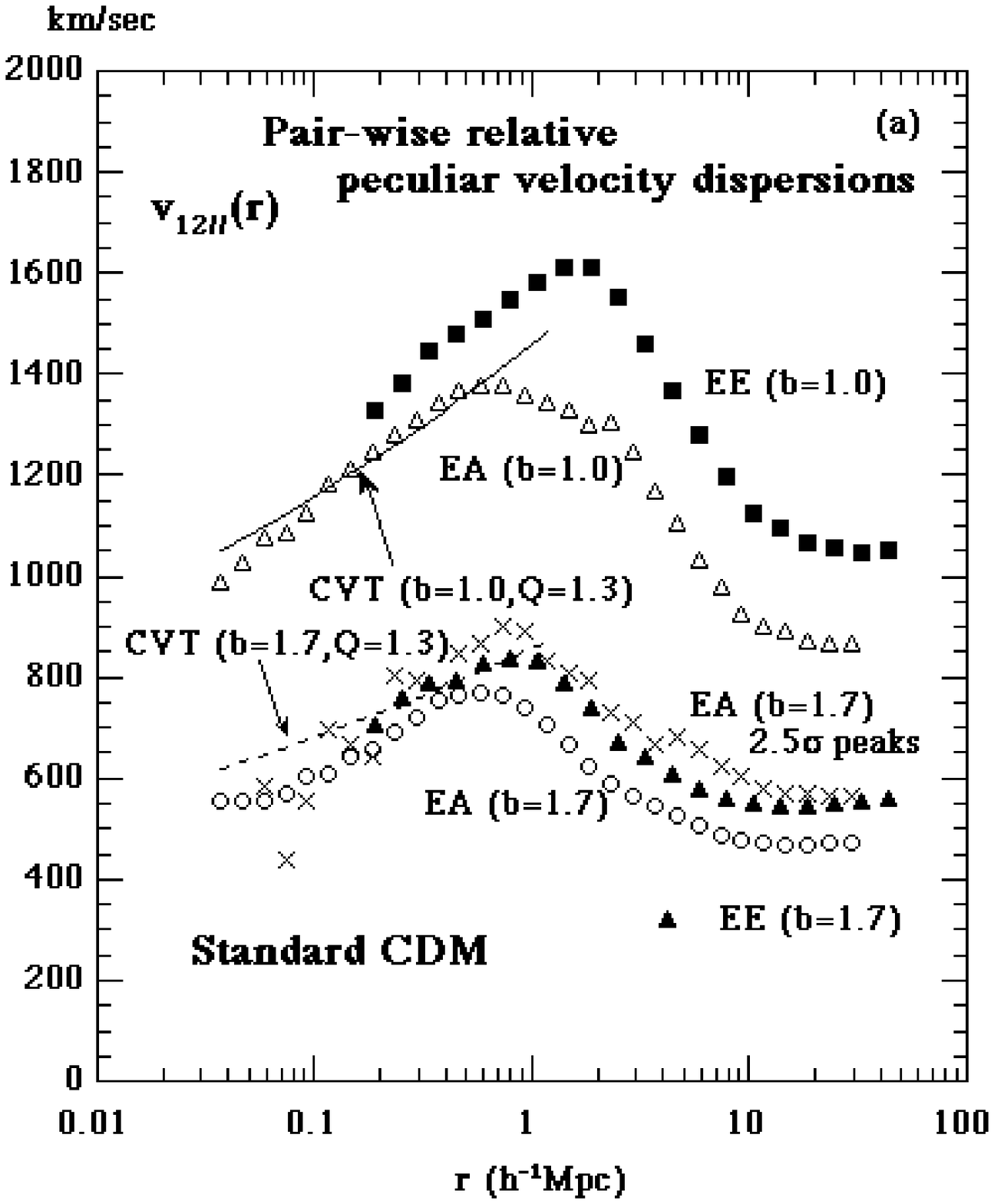,height=9cm}
   \leavevmode\psfig{figure=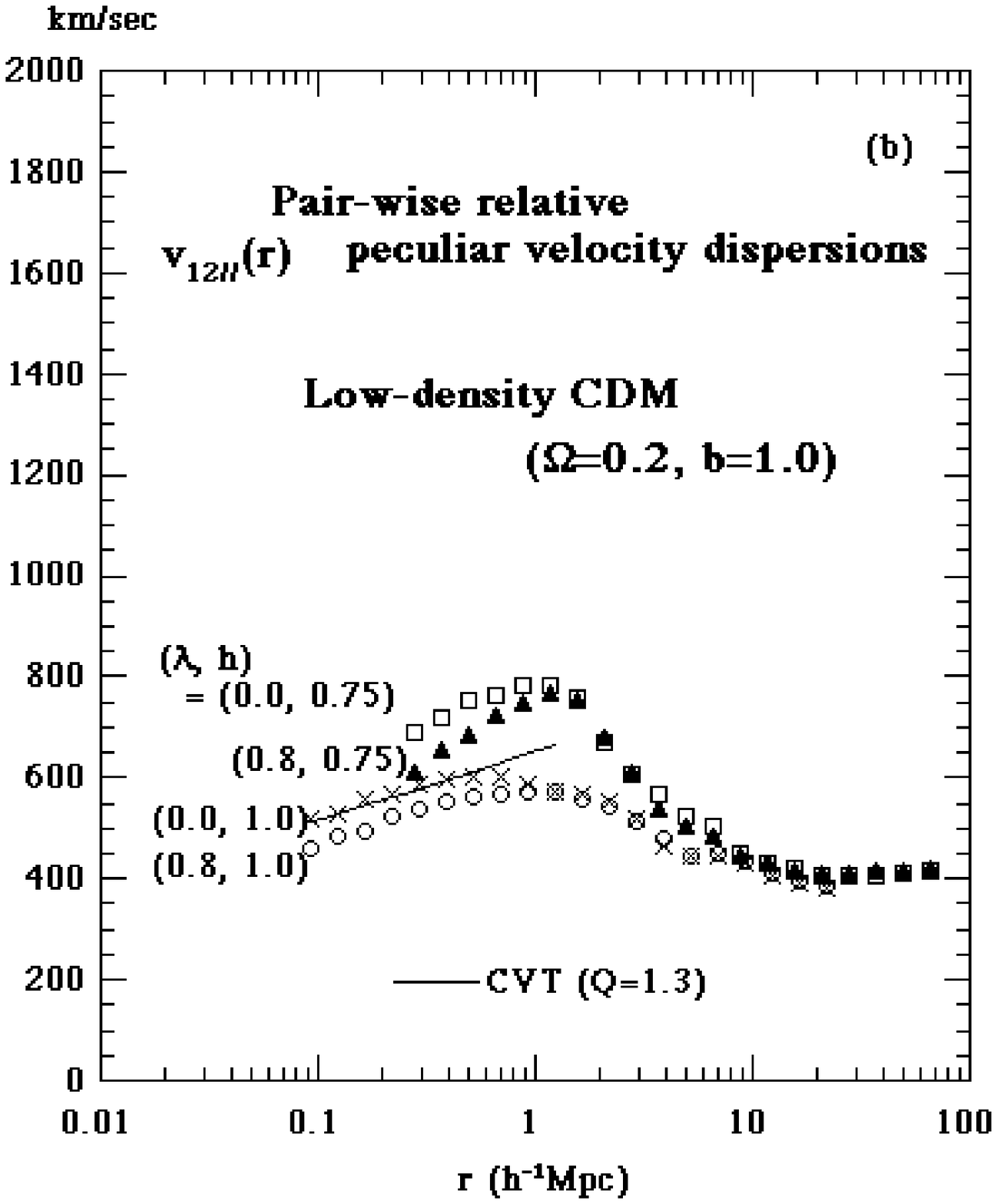,height=9cm}
\end{center}
\vspace{-0.5cm}
\caption{Pairwise relative peculiar velocity dispersions:
  comparison between predictions of the CVT and simulations (Suto
  1993).  (a) Standard CDM ($\Omega_0=1$, $\lambda_0=0$, $h=0.5$):
  Model EA employs $N=128^3$ particles in comoving $(130\himpc)^3$
  box, while model EE employs $N=128^3$ particles in comoving
  $(200\himpc)^3$ box. The different symbols denote the results at
  different epochs for the models; EE at $b_{\rm g}=1$ ({\it filled
    squares}), EA at $b_{\rm g}=1$ ({\it open triangles}), EE at
  $b_{\rm g}=1.7$ ({\it filled triangles}), EA at $b_{\rm g}=1.7$
  ({\it open circles}), and EA at $b_{\rm g}=1.7$ but for $2.5\sigma$
  peak particles ({\it crosses}).  The solid and dashed lines show the
  CVT prediction $\sigma_{\oned, \CVT}(r)$ for $b_{\rm g}=1.0$ and
  $1.7$, respectively ($Q_\rho=1.3$).  (b) Low-density CDM
  ($\Omega_0=0.2$, $b=1.0$): $(\lambda_0,h) =(0,0.75)$ model with
  $N=128^3$ particles in comoving $(300\himpc)^3$ box ({\it open
    squares}), $(0.8,0.75)$ model with $N=128^3$ particles in comoving
  $(300\himpc)^3$ box ({\it filled triangle}), $(0.0,1.0)$ model with
  $64^3$ particles in comoving $(100\himpc)^3$ box ({\it crosses}),
  and $(0.8,1.0)$ model with $N=64^3$ particles in comoving
  $(100\himpc)^3$ box ({\it open circles}).
\label{fig:ptpcvt}
}
\end{figure}

To what extent is the CVT prediction reliable ?  In order to examine
this, I compared $\sigma_{\oned, \CVT}(r)$ with the one-dimensional
peculiar velocity dispersions of particle pairs with separation $r =
|{\mb r}_1 - {\mb r}_2 |$ directly computed from N-body simulations
for cold dark matter models: $v_{12 \sVert} (r) \equiv \langle [ ({\mb
  v}_1 - {\mb v}_2 ) \cdot ({\mb r}_1 - {\mb r}_2 )/|{\mb r}_1 - {\mb
  r}_2 | ]^2 \rangle^{1/2} $. Figure \ref{fig:ptpcvt} summarizes the
comparison which implies that $\sigma_{\oned, \CVT}(r)$ reproduces the
simulation result excellently for $0.1\himpc \simlt r \simlt 1\himpc$;
CVT is quite reliable in predicting $v_{12 \sVert}(r)$ for $r <
1\himpc$. It should be stressed that the crucial assumption in
deriving the prediction (\ref{eq:cvt1d}) is equation (\ref{eq:hier}),
and the result is independent of the theoretical model for dark
matter. In this sense the prediction (\ref{eq:cvt1d}) is general, and
the good agreement in CDM models should be ascribed to the fact that
the CDM models actually satisfy the relation (\ref{eq:cvt1d}).

Using the anisotropy of the two-point correlation function of the CfA1
galaxy redshift survey, Davis and Peebles\cite{dp83} estimated the
line of sight peculiar velocity dispersions $\sigma_{\rm g,obs}^2(r_p)$
of galaxy pairs seen projected at separation $r_p$. This is related to
the dispersions $ \langle v^2_{\threed,21} (r) \rangle$ for galaxy
pairs with separation $r$ in three dimension as
\begin{equation}
\sigma_{\rm g,obs}^2(r_p) = 
{\displaystyle
\int_0^\infty dy\, \xi_{\rm g}\left(\sqrt{r_p^2+y^2}\right) \,
\left\langle v^2_{\threed,21} \left(\sqrt{r_p^2+y^2}\right) \right\rangle
\over
\displaystyle 3 \int_0^\infty dy\, \xi_{\rm g}\left(\sqrt{r_p^2+y^2}\right)} . 
\label{eq:sigmarp}
\end{equation}
Note that Davis and Peebles\cite{dp83} found that $\sigma_{\rm
  g,obs}^2(r_p) \propto r_p^{0.13\pm0.04}$ which justifies our
assumption of the {\it mass} correlation function $\xi(r) \propto
r^{-1.8}$ as $\xi_{\rm g}(r)$ from a dynamical point of view.  If
$\xi(r) \propto r^{-\gamma}$ and $\langle v^2_{\threed,21} (r) \rangle
\propto r^{2-\gamma}$, the corresponding CVT estimator corrected for
the projection becomes
\begin{equation}
\sigma_{\oned, \CVT, proj}(r_p) 
= C(\gamma) \,   \sigma_{\oned, \CVT}(r_p)
\qquad
C(\gamma) \equiv \sqrt{\Gamma(\gamma/2) \, \Gamma(\gamma - 3/2)
 \over  \Gamma(\gamma/2 - 1/2) \, \Gamma(\gamma -1)} ,
\label{eq:cvtproj}
\end{equation}
where $\Gamma(x)$ is the Gamma function.  Thus $C(\gamma)$ is the
correction factor for the projection effect\cite{suto93,dp83};
$C(1.65)\sim 1.36$, $C(1.8)\sim 1.11$, $C(1.95)\sim 1.02$.

Independent estimates of $\sigma_{\rm g,obs}(r_p)$ are available from
several galaxy redshift
catalogues\cite{dp83,sdp,spn,mo,fisher,marzke,guzzo}.  As pointed out
earlier by Mo, Jing, and B\"{o}rner\cite{mo}, the currently available
redshift catalogues are far from the fair sample, and the
observational estimate of $\sigma_{\rm g,obs}^2(r_p)$ may differ from
its true cosmic average due to the limited survey volume and the
selection effect.  Therefore it is meaningful to re-examine the
problem in more details.

If the hierarchical relation (\ref{eq:hier}) holds as is indicated
from the galaxy distribution, the amplitudes of the two- and
three-point correlation functions, or equivalently $b_{\rm g}$ and
$Q_\rho$, are the two uncertain parameters in the CVT prediction
(\ref{eq:cvt1d}). Recent numerical and analytical studies in nonlinear
gravitational clustering seem to indicate that $Q_\rho$ can be
approximated as a constant in the range of $0.5$ and $2$ which depends
very weakly on the underlying cosmological model.  Thus if the value
of $b_{\rm g}$ is fixed from the COBE data assuming CDM models, for
example, the CVT prediction (\ref{eq:cvt1d}) is completely specified.

\begin{figure}
\vspace*{-2.0cm}
\begin{center}
   \leavevmode\psfig{figure=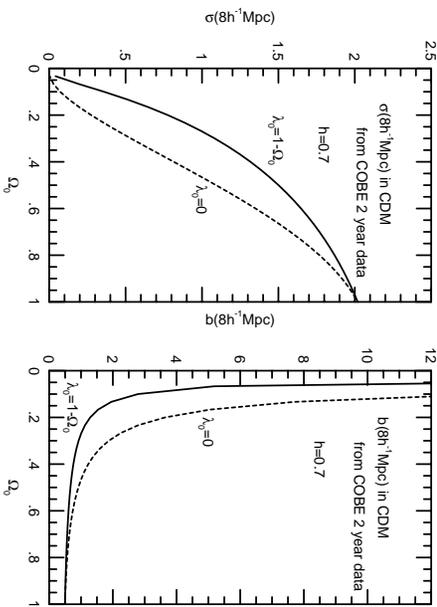,height=8cm,angle=180}
\end{center}
\caption{CDM spectrum normalization factor from COBE 2 year data.
  (a) top-hat mass rms fluctuation, $\sigma(8\himpc)$; (b) the biasing
  factor $b(8\himpc) \equiv 1/\sigma(8\himpc)$ in
  $\lambda_0=1-\Omega_0$ (solid curve) and $\lambda_0=0$ (dashed
  curve) CDM models ($h=0.7$ is assumed).
\label{fig:cobesigma8}
}
\end{figure}

Figure \ref{fig:cobesigma8} plots the empirical fit for the COBE
normalized $\sigma(8\himpc)$ and $b(8\himpc) \equiv 1/\sigma(8\himpc)$
by Nakamura\cite{nakamura} to the numerical computation by
Sugiyama\cite{sugiyama}.  There I adopt the baryon density parameter
$\Omega_b=0.015h^{-2}$.  Incidentally it is amusing to note that
$\Omega_0=0.2$, $\lambda_0=0.8$, and $h=0.7$ CDM model just
corresponds to $b_{\rm g}=1$, i.e., {\it galaxies faithfully trace
  mass} in this model according to the COBE normalization.

\begin{figure}
\vspace{-3.0cm}
\begin{center}
\hspace{1cm} \leavevmode\psfig{figure=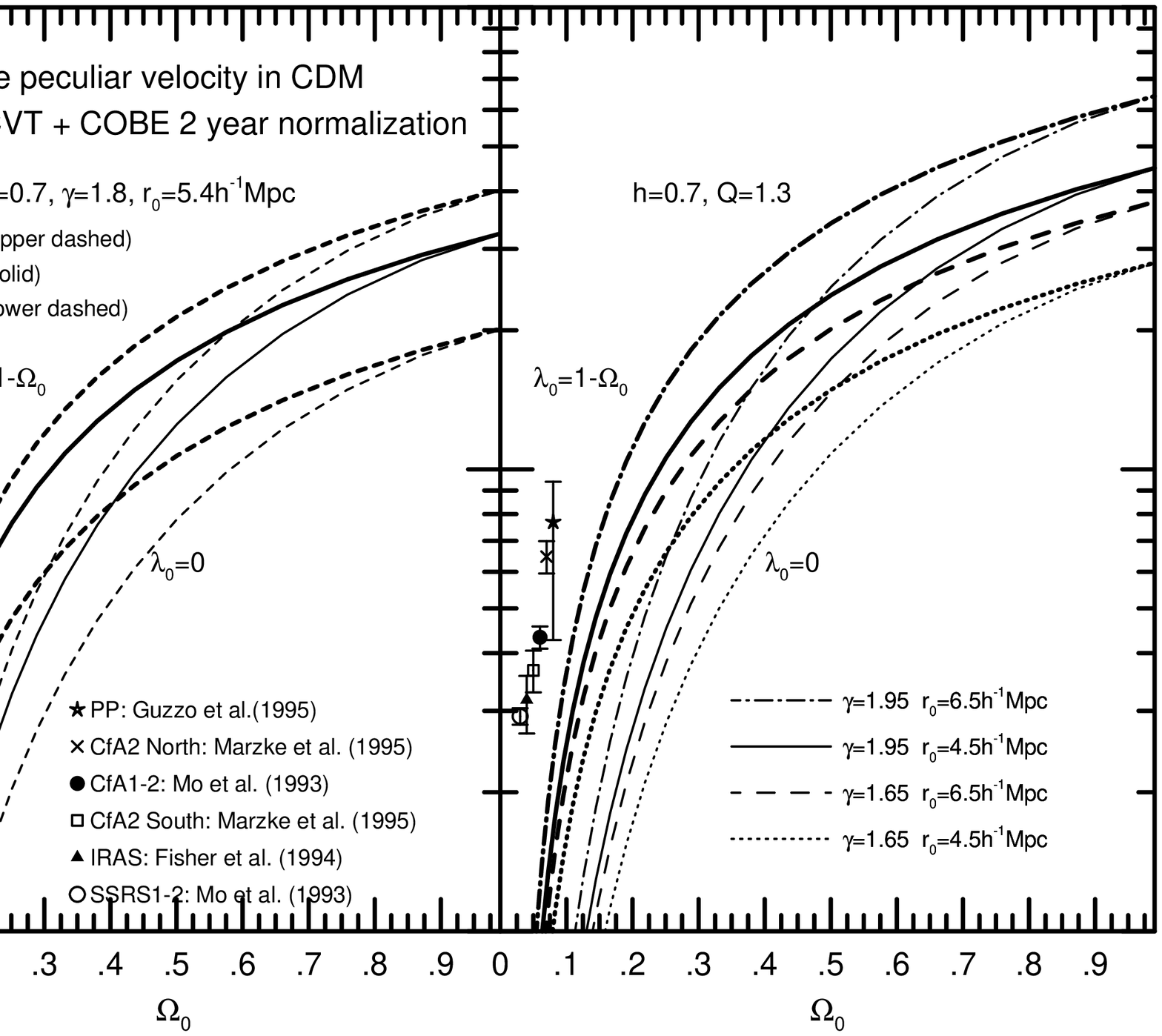,height=10cm,angle=180}
\vspace{-0.1cm}
\end{center}
\caption{Pairwise relative peculiar velocity dispersions
  at $1\himpc$ from the observations (symbols) compared with the
  predictions of the cosmic virial theorem in $\lambda_0=1-\Omega_0$
  (thick curves) and $\lambda_0=0$ (thin curves) CDM models.
\label{fig:sigma12rp}
}
\end{figure}

Once the $b_{\rm g}$ is specified, it is easy to compute the CVT
prediction (\ref{eq:cvt1d}).  Figure \ref{fig:sigma12rp} plots
$\sigma_{\oned, \CVT, proj}(r=1\himpc)$ (eq.[\ref{eq:cvtproj}]) as a
function of $\Omega_0$. The symbols denote several recent
observational estimates which are shifted along the x-axis just for
illustration. The left panel shows the dependence on $Q_\rho$ and the
right panel shows how the CVT prediction is sensitive to their {\it
  local} value, or sample-to-sample variation of the data.

Figure \ref{fig:sigma12rp} exhibits that both observational
estimates and CVT predictions of $\sigma_{\oned}(r=1\himpc)$ vary by a
factor of $2\sim3$ depending on the local clustering degree of each
sample. Even allowing for these, however, it looks unlikely that
high-density CDM models ($\Omega_0 \simgt 0.5$) are compatible with
the currently observed values of $\sigma_{\oned}(r=1\himpc)$.  In a
similar argument, one may obtain strong upper limits on $\Omega_0$ in
general cosmological scenarios as long as $b_{\rm g}$ is not much
greater than unity.

\section{$\Omega_0$ and $\lambda_0$ from quasar clustering at high redshifts}

As shown in the previous section, small-scale velocity dispersions of
galaxies place strong upper limits on $\Omega_0$, fairly independently
of $\lambda_0$. In turn, detailed analysis of clustering of objects at
high redshifts will place a potentially important constraint on
$\lambda_0$ via an effect which we call the {\it cosmological redshift
  distortion}\cite{ms96}. A very similar idea was put forward
independently by Ballinger, Peacock and Heavens\cite{bph} although
they work entirely in k-space. The result that I describe below
considers the analysis of two-point correlation functions, which would
be more straightforward to derived from the quasar catalogues.

Let us consider a pair of objects located at redshifts $z_1$ and $z_2$
whose redshift difference $\delta z \equiv z_1-z_2$ is much less than
the mean redshift $z \equiv (z_1+z_2)/2$.  Then the observable
separations of the pair parallel and perpendicular to the
line-of-sight direction, $s_{\sVert}$ and $s_\bot$, are given as
$\delta z/H_0$ and $z\delta\theta/H_0$, respectively, where $H_0$ is
the Hubble constant and $\delta\theta$ denotes the angular separation
of the pair on the sky.  The cosmological redshift-space distortion
originates from the anisotropic mapping between the redshift-space
coordinates, $(s_\sVert, s_\bot)$, and the real comoving
ones\cite{ms96}, $(x_\sVert, x_\bot) \equiv (c_\sVert s_\sVert, c_\bot
s_\bot)$; $c_\bot$ is written in terms of the angular diameter
distance $D_\A$ as $c_\bot = H_0 (1+z) D_\A/z$, and
\begin{equation}
\label{eq:csvert}
c_\sVert (z) = {H_0 \over H(z)} = {1 \over \sqrt{\Omega_0 (1 + z)^3 +
    (1-\Omega_0-\lambda_0) (1 + z)^2 + \lambda_0} } .
\end{equation}

The relation between the two-point correlation functions of quasars in
redshift space, $\xi^{(s)}(s_\bot, s_\sVert)$, and that of {\it mass}
in real space $\xi^{(r)}(x)$ can be derived in linear
theory\cite{hamilton,ms96}:
\begin{eqnarray} 
\label{eq:xis}
   &&
   \xi^{(s)}(s_\bot, s_\sVert) = 
   \left(1+{2\over 3}\beta(z) +{1\over5}[\beta(z)]^2\right) \xi_0(x)
   P_0(\mu) 
   -\left({4\over 3}\beta(z)
   +{4\over7}[\beta(z)]^2\right)\xi_2(x) P_2(\mu)
   \nonumber\\
   && \qquad\qquad
   +\frac{8}{35}[\beta(z)]^2 \xi_4(x) P_4(\mu) ,
\end{eqnarray}
where $x \equiv \sqrt{{c_\sVert}^2 {s_\sVert}^2 + {c_\bot}^2
  {s_\bot}^2}$, $\mu\equiv c_{\sVert} s_{\sVert} /x$, $P_n$'s are the
Legendre polynomials, 
\begin{eqnarray} 
\label{eq:xi2l}
\beta(z) \equiv {1 \over b(z)}\frac{d\ln D(z)}{d\ln a}, \qquad
\xi_{2l}(x) 
= { (-1)^l \over x^{2l+1}} \left(\int_0^x xdx\right)^l x^{2l}
\left({d \over dx}{1 \over x}\right)^l x \xi^{(r)}(x),
\end{eqnarray}
and $D(z)$ is the linear growth rate.  

\begin{figure}
\vspace*{-4.0cm}
\begin{center}
   \leavevmode\psfig{figure=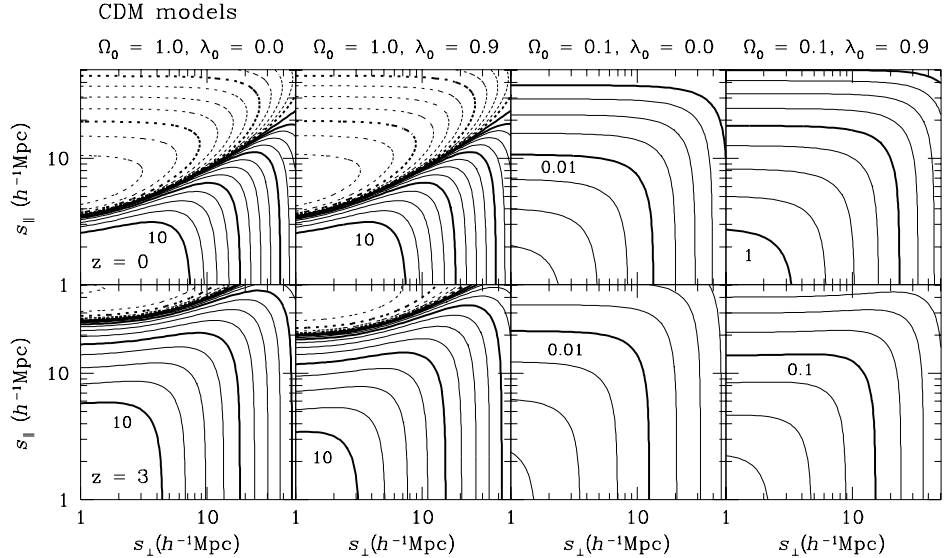,width=15cm}
\end{center}
\vspace{-3.0cm}
\caption{
The contours of $\xi^{(s)}(s_\bot,
  s_{\sVert})$ in CDM models at $z=0$ ({\it upper panels}) and $z=3$
  ({\it lower panels}).  
  Solid and dashed lines indicate
  the positive and negative $\xi^{(s)}$, respectively.  Contour
  spacings are $\Delta {\rm log}_{10} |\xi| = 0.25$.
\label{fig:xicont}}
\end{figure}

For specific examples, we compute $\xi^{(s)}(s_\bot, s_{\sVert})$ in
linear theory applying equations (\ref{eq:xis}) and (\ref{eq:xi2l}) in
CDM models with $H_0=70$
km$\cdot$s$^{-1}\cdot$Mpc$^{-1}$\cite{tanvir}. The resulting contours
are plotted in Figure \ref{fig:xicont}. The four sets of values of
$\Omega_0$ and $\lambda_0$ are indicated at the top of each panel.  We
adopt the COBE normalization\cite{nakamura,sugiyama,ws}.

\begin{figure}
\vspace*{-1.0cm}
\begin{center}
   \leavevmode\psfig{figure=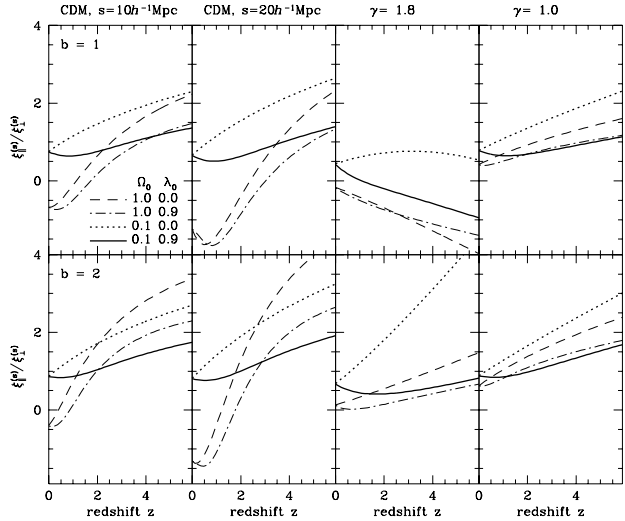,width=10cm}
\end{center}
\vspace{-2.0cm}
\caption{The anisotropy parameter
  $\xi^{(s)}_{\sVert}(s)/\xi^{(s)}_\bot(s)$ as a function of $z$.
  Upper and lower panels assume that $b=1$ and $2$, respectively. From
  left to right, the panel corresponds to CDM at $s=10\himpc$, CDM at
  $s=20\himpc$, a power-law model with $\gamma=1.8$ and a power-law
  model with $\gamma=1.0$.
\label{fig:aniso}}
\end{figure}

In order to quantify the cosmological redshift distortion in Figure
\ref{fig:xicont}, let us introduce the anisotropy parameter
$\xi^{(s)}_{\sVert}(s)/\xi^{(s)}_\bot(s)$, where
$\xi^{(s)}_\bot(s) \equiv \xi^{(s)}(s,0)$ and
$\xi^{(s)}_{\sVert}(s) \equiv \xi^{(s)}(0,s)$. The left
four panels in Figure \ref{fig:aniso} show the anisotropy parameter
against $z$ in CDM models. This clearly exhibits the extent to which
one can discriminate the different $\lambda_0$ models on the basis of
the anisotropies in $\xi^{(s)}$ at high redshifts.

\section{Conclusions}

One can roughly divide the source of cosmological information into
four regimes; (i) linear regime ($r\simgt 10\himpc$) at $z\sim 0$,
(ii) nonlinear regime ($r\sim 1\himpc$) at $z\sim 0$, (iii) linear
regime at $z\gg 1$, and (iv) nonlinear regime at $z\gg 1$.

The clustering feature at $z\sim 0$ is best probed by galaxy redshift
surveys, and the current samples, albeit statistically limited by the
number of galaxies, are heavily used for a variety of cosmological
tests. The conventional redshift-space distortion
analysis\cite{kaiser,hamilton} in the regime (i) yields generally
$\Omega_0 \ll 1$; $\Omega_0^{0.6}/b_{\rm g}(z=0) = 0.55 \pm 0.12$ for
instance from the Durham/UKST galaxy redshift survey of $\sim 2500$
galaxies.  I have shown in the first part of the talk that the
small-scale peculiar velocity dispersions of galaxies in comparison
withe the CVT prediction constrains the value of $\sqrt{\Omega_0
  Q_\rho}/b_{\rm g}(z=0)$.  Allowing for the sample-to-sample
variation of the available redshift surveys and the uncertainties for
$Q_\rho$ and $b_{\rm g}(z=0)$, I still conclude that the observation
in the regime (ii) favors low-density universes $\Omega_0 \simlt 0.5$
at most.

In contrast to the regimes (i) and (ii), the proper analysis of the
regimes (iii) and (iv) requires a large number of quasars
homogeneously samples which are not currently available. Therefore the
cosmological tests in theses regimes have not been fully explored even
theoretically. In the second part of the talk, I presented an example
in this line of investigations, cosmological redshift-space
distortion\cite{ms96}. I derived the formula which describes the
degree of the anisotropies of two-point correlation functions of
quasars in linear regime (iii), and argued that it is a potentially
powerful discriminator of $\lambda_0$ once $\Omega_0$ is determined
from the nearby observation. From the observational point of view,
however, it would be much easier to detect the the anisotropies of
two-point correlation functions in the regime (iv).  We are currently
approaching this important area of research using the nonlinear theory
and simulations\cite{sm96,mms}.

I do look forward to the next-generation redshift surveys which will
definitely provide data catalogues and thus make feasible these
precise cosmological tests.

\bigskip

\acknowledgments

The second part of the present talk is based on my collaborative work
with Takahiko Matsubara. I thank him for the fruitful and enjoyable
collaboration, and Yi-Peng Jing, Takahiro T. Nakamura and David
Weinberg for discussions. I am grateful to John Peacock for calling my
attention to their work\cite{bph} prior to publication after we
submitted the paper\cite{ms96} of the second part of the present talk.

This research was supported in part by the Grant-in-Aid by the
Ministry of Education, Science, Sports and Culture of Japan (07CE2002)
to RESCEU (Research Center for the Early Universe), the University of
Tokyo.  I thank the session organizers, Chul Hoon Lee and Katsuhiko
Sato, for inviting me to give the present talk at Gravitation and
Cosmology session of the APCTP inauguration conference.


\begin{references}

\vspace*{-1cm}
\baselineskip=13pt
\parskip1pt

\bibitem{suto93} Y. Suto, Prog.Theor.Phys. {\bf 90}, 1173 (1993).

\bibitem{ms96} T. Matsubara and Y. Suto, Astrophys.J. (Letters),
  October 10 issue, in press (1996).

\bibitem{peebcvt} P.J.E. Peebles, Astrophys.Sp.Sci.,{\bf 45},3 (1976).

\bibitem{lss} P.J.E. Peebles,The Large Scale Structure of the
  Universe, Princeton University Press (1980).

\bibitem{gp77} E.J. Groth and  P.J.E. Peebles, Astrophys.J. {\bf 217},
  385 (1977).

\bibitem{dp83} M. Davis and  P.J.E. Peebles, Astrophys.J. {\bf 267},
  465 (1983).

\bibitem{ms94} T. Matsubara and Y. Suto, Astrophys.J. {\bf 420} 497
  (1994).

\bibitem{sm94} Y. Suto and T. Matsubara, 
Astrophys.J. {\bf 420}, 504 (1994).  

\bibitem{uis} H. Ueda, M. Itoh, and Y. Suto, Astrophys.J. {\bf 408}, 3
  (1993).

\bibitem{sdp} R.S. Somerville, M.Davis, and J.R. Primack, preprint
    astro-ph/9604041.

\bibitem{spn} R.S. Somerville, J.R. Primack, and R. Nolthenius,
    preprint astro-ph/9604051.

\bibitem{mo} H.J. Mo, Y.P. Jing, and G. B\"{o}rner,
  Mon.Not.R.Astron.Soc., {\bf 264}, 825(1993).

\bibitem{fisher} K.B.Fisher, M.Davis, M.A.Strauss, A.Yahil, and J.Huchra,
  Mon.Not.R.Astron.Soc., {\bf 267}, 927(1994).

\bibitem{marzke} R.O.Marzke, M.J.Geller, L.N. da Costa, and J.Huchra,
  Astron.J., {\bf 110}, 477(1995).

\bibitem{guzzo} L.Guzzo, K.B.Fisher, M.A.Strauss, R. Giovanelli, and
  M.P.Haynes, astro-ph/9503114, Astrophys.Lett. and Communications, in 
  press.

\bibitem{nakamura} T. T. Nakamura, master thesis to the University of
  Tokyo, unpublished (1996).

\bibitem{sugiyama} N. Sugiyama, Astrophys.J.Suppl. {\bf 100}, 281(1995).

\bibitem{tanvir} N.R. Tanvir, T. Shanks, H.C. Ferguson, 
and D.R.T. Robinson, Nature, {\bf 377} 27 (1995).

\bibitem{bph} W. E. Ballinger, J. A. Peacock and A. F. Heavens,
  Mon.Not.R.Astron.Soc., (1996), in press.

\bibitem{ap} C. Alcock and B. Paczy\'nski, Nature {\bf 281}, 358
  (1979).

\bibitem{phil} S. Phillipps, Mon.Not.R.Astron.Soc. {\bf 269}
  1077(1994).

\bibitem{ryden} B. Ryden,  Astrophys.J. {\bf 452} 25 (1995).

\bibitem{ppc} P.J.E. Peebles, Principles of Physical Cosmology,
  Princeton University Press (1993).

\bibitem{lahav} O. Lahav,  P.B. Lilje, J.R. Primack, and M.J. Rees,
Mon.Not.R.Astron.Soc. {\bf 251} 128(1991).

\bibitem{kaiser} N. Kaiser, Mon.Not.R.Astron.Soc. {\bf 227} 1(1987).

\bibitem{hamilton} A.J.S. Hamilton, Astrophys.J. {\bf 385} L5 (1992).

\bibitem{ws} M. White and D. Scott, Astrophys.J. 
{\bf 459} 415 (1996).

\bibitem{ratcliffe} A. Ratcliffe et al., preprint astro-ph/9602062.

\bibitem{sm96} Y. Suto and T. Matsubara, submitted to Astrophys.J.  

\bibitem{mms} H. Magira, T. Matsubara and Y. Suto, in preparation.

\end{references}
\end{document}